\documentclass[10pt]{article}
\pdfoutput=1
\usepackage{setspace}
\usepackage{psfrag}
\usepackage{latexsym}
\usepackage{indentfirst}
\usepackage{fancyhdr}
\usepackage{amssymb}
\usepackage{amsmath}
\usepackage{amsfonts}
\usepackage{pifont}
\usepackage{bbold}
\usepackage{color}
\usepackage{endnotes}

\usepackage{graphicx}
\usepackage[center,footnotesize,hang]{subfigure}
\usepackage{array}

\DeclareMathAlphabet{\mathsc}{OT1}{cmr}{m}{sc}
\newcommand {\ignore}[1]{}

\def\lnv{lepton number violation }
\def\10{$SO(10)$}
\def\21{SU(2) $\otimes$ U(1) }

\def\422{$SU(4) \otimes SU(2) \otimes SU(2)$}
\def\321{SU(3) $\otimes$ SU(2) $\otimes$ U(1)}
\def\gsim{\raise0.3ex\hbox{$\;>$\kern-0.75em\raise-1.1ex\hbox{$\sim\;$}}}
\def\lsim{\raise0.3ex\hbox{$\;<$\kern-0.75em\raise-1.1ex\hbox{$\sim\;$}}}
\def\lsim{\raise0.3ex\hbox{$\;<$\kern-0.75em\raise-1.1ex\hbox{$\sim\;$}}}
\def\gsim{\raise0.3ex\hbox{$\;>$\kern-0.75em\raise-1.1ex\hbox{$\sim\;$}}}
\def\vev#1{\left\langle #1\right\rangle}
\def \znbb {$0\nu\beta\beta$ }
\def \jznbb {$J0\nu\beta\beta$ }
\def\SM{$\mathrm{SU(3)_c \otimes SU(2)_L \otimes U(1)_Y}$ }

\newcommand{\AddrAHEP}{%
  AHEP Group, Institut de F\'{\i}sica Corpuscular --
  C.S.I.C./Universitat de Val{\`e}ncia,  Valencia, Spain}

\newcommand{\Cinvestav}{Departamento de F\'{\i}sica, Centro de
  Investigaci{\'o}n y de Estudios Avanzados del IPN M\'exico, DF, Mexico}
 
\renewcommand{\Large}{\large}

\newcommand{\ba}{\begin{array}}
\newcommand{\ea}{\end{array}}
\relax
\def\321{$SU(3)\times SU(2)\times U(1)$}

\begin{document}

\noindent{\bf \Large Invisible decays of ultra-high energy neutrinos} \\ 

\noindent
L. Dorame$^a$\footnote{dorame@ific.uv.es}, O. G. Miranda$^b$\footnote{omr@fis.cinvestav.mx}, 
J. W. F. Valle$^a$\footnote{valle@ific.uv.es} \\
$^a${\it \AddrAHEP} \\
$^b${\it \Cinvestav} \\


\noindent {\bf Abstract} \\
\noindent {\it 
Gamma-ray bursts (GRBs) are expected to provide a source of ultra
high energy cosmic rays, accompanied with potentially detectable
neutrinos at neutrino telescopes.  Recently, IceCube has set an
upper bound on this neutrino flux well below theoretical
expectation. We investigate whether this mismatch between
expectation and observation can be due to neutrino decay. We
demonstrate the phenomenological consistency and theoretical  
plausibility of the neutrino decay hypothesis. A potential
implication is the observability of majoron-emitting neutrinoless
double beta decay. }

\vspace*{1cm}

The source of ultra high energy cosmic rays remains a mystery.  In
gamma-ray burst (GRB) models such as the fireball model cosmic-ray
acceleration should be accompanied by neutrinos produced in the decay
of charged pions created in interactions between the high-energy
protons and $\gamma$-rays~\cite{Waxman:1997ti}. Recently the Ice-Cube
collaboration reported an upper limit on the flux of energetic
neutrinos associated with GRBs almost four times below this
prediction~\cite{Abbasi:2012zw}.

Various possible explanations have been considered to explain the non
observation of this ultra high energy neutrino flux. For example, a
complete detailed numerical analysis of the fireball neutrino model
predicts a neutrino flux that is one order of magnitude lower than the
analytical computations~\cite{Hummer:2011ms}.  On the other hand,
another recent computation~\cite{He:2012tq} of the neutrino flux in
the fireball model gives a mild reduction in the neutrino flux if a
relation between the bulk Lorentz factor, $\Gamma$, and the burst
energy is assumed.  Finally, based on the specific case of GRB
130427A, it has been argued that the low neutrino flux can be
explained with relatively large values for the bulk Lorentz factor
($\Gamma\ge 500$) and for the dissipation radius ($R_d\ge
10^{14}$~cm); it was shown in the same reference that both the
internal shock and the baryon photosphere models satisfied these
conditions.  

Here we focus on a different approach to explain the neutrino flux
deficit. Instead of studying the astrophysical mechanism of the source
objects, we look for a high energy physics explanation.  Some
mechanisms involving new physics in order to explain a possible
deficit in the observed neutrino flux have already been suggested. For
instance, the possibility of an oscillations involving a quasi Dirac
neutrino~\cite{Valle:1982yw} has been considered in
Ref.~\cite{Esmaili:2012ac}; a specific model for this case has been
studied in~\cite{Joshipura:2013yba} and the possibility of a
resonance effect has also been
discussed~\cite{Miranda:2013wla}. Another mechanism recently
discussed has been the case of a spin precession into a sterile
neutrino as a result of a nonzero neutrino magnetic
moment~\cite{Schechter:1981hw} and the strong magnetic fields
expected to be present in a GRB~\cite{Barranco:2012xj}.

Here we speculate on the plausibility of the neutrino decay hypothesis
as a possible explanation for the mismatch between observation and
expectation. The most attractive possibility involves invisible
decays, which have been considered theoretically since the
eighties~\cite{Schechter:1981cv,valle:1983ua,gelmini:1984ea,gelmini:1983pe,gonzalezgarcia:1988rw},
and recently revisited for the case of GRB neutrino
fluxes~\cite{Pakvasa:2012db}.  These decays arise in models with
spontaneous violation of ungauged lepton
number~\cite{chikashige:1980ui}, though typically
suppressed~\cite{Schechter:1981cv}.
A natural scenario to test neutrino stability are astrophysical
objects~\cite{bahcall:1986gq,Keranen:1999nf,beacom:2002cb,beacom:2002vi}.
In particular, limits on Majoron couplings from solar and supernova
neutrinos have been obtained in Refs.~\cite{Kachelriess:2000qc}. For
non astrophysical constraints, for example from \znbb searches,
see~\cite{Tomas:2001dh,Lessa:2007up,Gando:2012pj,Argyriades:2009ph}.
Moreover, as already mentioned, recent results from the Pierre Auger
Observatory (PAO)~\cite{Abreu:2011pf}, ANTARES~\cite{Biagi:2011kg}
and IceCube~\cite{Abbasi:2008hr} have placed strong constraints on
the neutrino flux coming from distant ultra high energy (UHE)
neutrino sources.

Here we explore the phenomenological plausibility and theoretical
consistency of the decay hypothesis within a class of low-scale \SM
seesaw schemes with spontaneous family-dependent lepton number violation.
We show that the required neutrino decay lifetime range hinted by the
non observation of UHE muon neutrinos is theoretically achievable for
the majoron-emitting neutrino decays and, moreover, consistent with
all existing phenomenological constraints.

\begin{figure}[!h]
\centering
\includegraphics[width=0.4\textwidth]{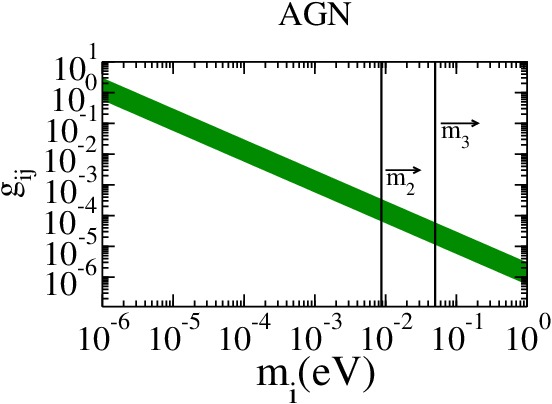}
\includegraphics[width=0.4\textwidth]{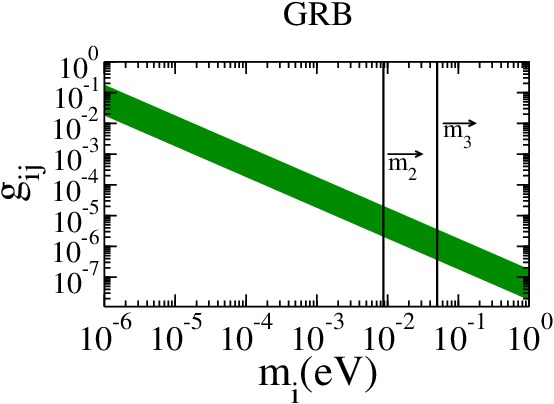}
\caption{Neutrino mass $m_i$ versus the required coupling constant
  $g_{ij}$ for the case of AGN's (left panel) and GRBs  (right panel).}
\label{fig:corr1}
\end{figure}

The decay rate $\nu_i \to \nu_j + J$ in the rest frame of $\nu_i$ is
\begin{equation}
\Gamma(\nu_i \to \nu_j + J) = \dfrac{g_{ij}^{2}}{16\pi}\dfrac{(m_i+m_j)^2}{m_i^3}(m_i^2-m_j^2) \ \ ,
\label{eq:decayrate}
\end{equation}
where $\nu_i$ and $\nu_j$ are active neutrinos and $J$ is a massless
or very light majoron associated to the spontaneous violation of
ungauged lepton number. Taking $m_j = 0$, we can estimate the decay
length (in meters) for a relativistic neutrino as given by
\begin{equation}
  L = c\tau = \dfrac{cE_i}{m_i \Gamma} \simeq 1 \times 10^9 \left(g_{ij}^{-2}\right) \left(\dfrac{E_i}{100 \mbox{TeV}}\right) \left(\dfrac{1\mbox{eV}}{m_i}\right)^2 \mbox{m} .
\end{equation}
For typical AGN distances we obtain the required values of $g_{ij}$
for a given neutrino mass $m_i$ which would cause decay before
reaching the detector. In Fig. (\ref{fig:corr1}) we took AGN distances
from $3.6$~Mpc (the distance to Centaurus A) up to $100$~Mpc. In the
bottom panel of the same Fig.~(\ref{fig:corr1}) we have plotted the
corresponding result for GRBs, at typical distances of $10$ - $10^3$
Mpc. The vertical lines correspond to the relevant region for $m_2$
and $m_3$ for the case $m_1=0$.  We have explicitly verified that, for
the GRB case, this approximation is in agreement with more detailed
estimates of neutrino lifetime ranges~\cite{Baerwald:2012kc}.

\begin{figure}[!h]
\centering
\includegraphics[width=0.43\textwidth]{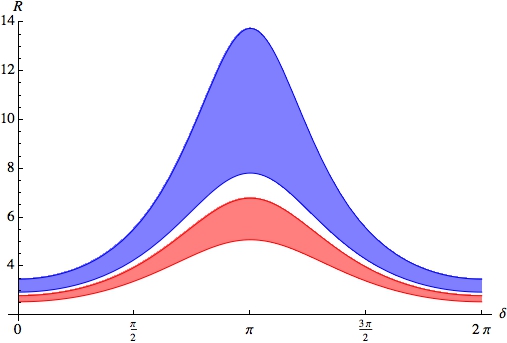}
\includegraphics[width=0.43\textwidth]{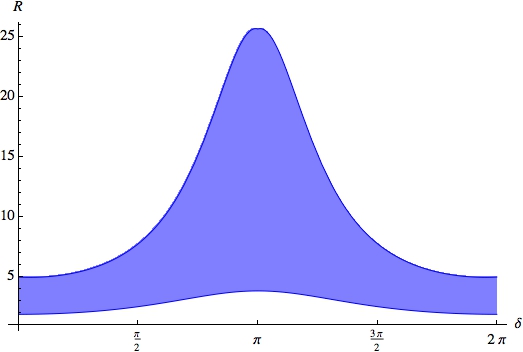} 
\caption{$R_{\nu_e:\nu_\mu}$ versus the CP phase $\delta$ for neutrino
  mixing angles~\cite{Tortola:2012te} at 1$\sigma$ (left panel) and
  at 3$\sigma$ (right panel). For the 1$\sigma$ case we show two
  regions: one for $\theta_{23}$ in the $1^{st}$ octant (orange lower
  band) and another one for $\theta_{23}$ in the $2^{nd}$ octant (blue
  higher band). One can see that values of $\theta_{23}$ in
  the $2^{nd}$ octant give a stronger effect. }
\label{fig:Mariam}
\end{figure}

As we will discuss below, putting into a theory context, such
couplings are fairly large to achieve theoretically.
In order to have an estimate of the neutrino flux reduction resulting
from neutrino decay we note that, since coherence is lost, the final
flux of a given neutrino flavour will be
\begin{equation}
\label{osc}
\phi_{\nu_\alpha}(E) = \sum_{i\beta}\phi_{\nu_\beta}^{\rm source}(E)|U_{\beta i}|^2|U_{\alpha i}|^2 e^{-L/\tau_i(E)} ,
\end{equation}
where $\phi$'s are neutrino fluxes at production and detection, L the
travel distance, $\tau_i$ the neutrino lifetime in the laboratory
frame and $U_{\alpha i}$ the elements of the lepton mixing
matrix~\cite{Schechter:1980gr}.  Typical neutrino energies lie in
the range of $10^5$~TeV and $10^3$~TeV for AGNs and GRBs respectively.
Note that, in the limit that $L\gg\tau_i$ where only the stable
state survives Eq.~(\ref{osc}) becomes
\begin{equation}
\label{osc1}
\phi_{\nu_\alpha}(E) = \sum_{i(\rm stable)\beta}\phi_{\nu_\beta}^{\rm source}(E)
|U_{\beta i}|^2|U_{\alpha i}|^2 .
\end{equation}
Here we take a normal hierarchy neutrino mass spectrum, the
disappearance of all states except the lightest (in this case $\nu_1$)
is allowed.  The final flux of $\nu_e$, $\nu_\mu$ and $\nu_\tau$ can
be computed from Eq. (\ref{osc1}) and will depend on the three mixing
angles and the Dirac CP phase $\delta$. In particular, we can
calculate the suppression of the muon neutrino flux, $\phi_{\nu_\mu}$,
using the ratio
\begin{equation}
R_{\nu_e:\nu_\mu}\equiv\dfrac{\phi_{\nu_e}}{\phi_{\nu_\mu}} =\left(\dfrac {\cos{\theta_{12}}\cos{\theta_{13}}}{|-\sin{\theta_{12}}\cos{\theta_{23}}-\sin{\theta_{13}}\sin{\theta_{23}}\cos{\theta_{12}} e^{\imath \delta}|}\right)^2 , 
\end{equation}
where $\theta_{12}$, $\theta_{23}$ and $\theta_{13}$ are the neutrino
mixing angles determined in neutrino oscillation experiments. 
The left panel of Fig.~(\ref{fig:Mariam}) shows the expected
values for this ratio when the neutrino mixing angles lie within the
$1\sigma$ bands from their current global best fit
values~\cite{Tortola:2012te,GonzalezGarcia:2012sz,Fogli:2012ua}
One can also see, in the right panel of 
the same 
Fig.~(\ref{fig:Mariam}), that by allowing these parameters to vary up
to their three sigma ranges, $R_{\nu_e:\nu_\mu}$ can be as large as 25
or as low as 2. Very similar results are found for global fit of
Ref.~\cite{GonzalezGarcia:2012sz}, as shown in Table~\ref{tab1}.

\begin{table}[h!]
\begin{center}
\begin{tabular}{|c||c|c|c|}
\hline
Global Fit&$1^{st}$octant&$2^{nd}$octant&At $3\sigma$\\
\hline
Forero, et. al.~\cite{Tortola:2012te}&2-7&3-14&2-25\\
\hline
Gonz\'alez-Garc\'{\i}a, et. al.~\cite{GonzalezGarcia:2012sz}&2-7&3-12&2-23\\
\hline
Fogli, et. al.~\cite{Fogli:2012ua}&2-6&-&2-22\\
\hline
\end{tabular}
\vskip .2cm
\caption{Allowed values of $R_{\nu_e:\nu_\mu}$ according to different neutrino oscillation fits. The second column shows the allowed region for the case of a 
$\theta_{23}$ in the first octant, while the third column shows the second 
octant case. Finally, the fourth column displays the allowed region if we 
consider the  3~$\sigma$ confidence level for $\theta_{23}$.}\label{tab1}
\end{center}
\end{table}

It is important to notice that, in this picture, the neutrino decay
will lead to a decrease in the muon neutrino flux while the electron
neutrino flux will increase. Alternatively in the presence of light
sterile states one can envisage a scenario where the muon neutrino
decays to the sterile state. Here we do not consider this case.
Recent
  data from Icecube reports the observation of two neutrinos with
  energies around $10^{15}$~eV, probably electron neutrinos~\cite{Aartsen:2013bka}.
  Moreover, there have been recent announcements of more neutrino
  events detected in IceCube~\cite{Whitehorn}.

%

We now turn to the issue of theoretical consistency of the decay
hypothesis.  
In most \SM seesaw models with spontaneous \lnv when one diagonalizes
the neutrino mass matrix one also diagonalizes, to first
approximation, the coupling of the resulting Nambu-Goldstone boson to
the mass eigenstate neutrinos~\cite{Schechter:1981cv}.
The exact form of the light-neutrino majoron couplings can be
determined explicitly by perturbative diagonalization of the seesaw
mass matrix, or by using a more general approach using only the
symmetry properties. The result is~\cite{Schechter:1981cv}
\begin{equation}\label{eq:Jnunu}
{g}_{ij} = -\frac{m_i}{v_1} \delta_{ij} +
\left[\frac{m_i}{v_1}\left(V_1^\dagger D^* {M}^{*-1}{M}^{-1}D^TV_1\right)_{ij}\right]_{\rm S} + \ldots
\end{equation}
where the subscript S denotes symmetrization, $D$ and $M$ are the
Dirac and Majorana mass terms in, say, the type-I seesaw scheme and
$V_1$ is the light neutrino diagonalization matrix.
One sees that the majoron couples proportionally to the light neutrino
mass, hence the coupling matrix is diagonal to first
approximation. The off-diagonal part of the $g_{ij}$ is inversely
proportional to three powers of \lnv scale $v_1\equiv \vev{\sigma}$,
since $M\propto v_1$. This is tiny, the only hope being to use a
seesaw scheme that allows for a very low \lnv scale, such as the
inverse seesaw~\cite{mohapatra:1986bd}.  The particle content is the
same as that of the Standard Model
(SM) except for the addition of a pair of two component gauge singlet
leptons, $\nu^c_i$ and $S_i$, within each of the three generations,
labeled by $i$. The isodoublet neutrinos $\nu_i$ and the fermion
singlets $S_i$ have the same lepton number, opposite with respect to
that of the three singlets $\nu^c_i$ associated to the
``right-handed'' neutrinos.
In the $\nu$, $\nu^c$, $S$ basis the $9\times9$ neutral lepton mass
matrix $M_{\nu}$ has the form:
\begin{equation}
M_{\nu}= \left[\begin{array}{ccc}
       0& m_D^T& 0\\
       m_D& 0& M^T\\
      0& M& \mu
      \end{array}
 \right],\label{Mnu}
\end{equation}
where $m_{D}\propto \vev{\Phi}$ is the standard Dirac term coming from
the SM Higgs vev and $M$ is a bare mass term. The term $\mu \propto
\vev{\sigma}$, the vacuum expectation value of $\sigma$ responsible for
spontaneous low-scale \lnv as proposed
in~\cite{gonzalezgarcia:1988rw}. This gives rise to a majoron J,
\begin{equation}
J=\sqrt{2} \ \mbox{Im}\sigma . 
\end{equation}

As a result of diagonalization one obtains an effective light
neutrino mass matrix. 
Note that lepton number symmetry is recovered as $\mu \to$ 0,
making the three light neutrinos strictly massless.
The majoron couplings of the light mass eigenstate neutrinos are
determined again as a sum of two pieces as in Eq.~(\ref{eq:Jnunu}).
Detailed calculation shows that its off-diagonal part behaves as $ P'
\sim \mu^2 D^2 {M}^{-4}$.
Even if the $M$ can be significantly lower than that of the standard
high-scale type-I seesaw it is clear that this is way too small in
order to produce neutrino decay within the relevant astrophysical
scales.

The only way out is to induce a mismatch between the neutrino mass
basis and the coupling basis. This can be achieved by making lepton
number a family-dependent
symmetry~\cite{gelmini:1984ea,gelmini:1983pe}. The model is by no
means unique, here we give an example based on \SM $\otimes U(1)_H$
assigned as shown in Table~\ref{tab2}

\begin{table}[h]
\begin{center}
\begin{tabular}{|c||c|c|c|c|c|c|c|c|c|c|c|}
\hline
&$\overline{L_e}$&$\overline{L_\mu}$&$\overline{L_\tau}$&$\nu_{R_e}$&$\nu_{R_\mu}$&$\nu_{R_\tau}$&$h$&$S_1$&$S_2$&$S_3$&$\sigma$\\
\hline
\hline
$SU(2)$&2&2&2&1&1&1&2&1&1&1&1\\
\hline
$U(1)_H$&$-2$&$-2$&$-4$&$2$&$2$&4&0&$0$&$0$&$-2$&$-2$\\
\hline
\end{tabular}
\vskip .2cm
\caption{Model field representation content and transformation properties}
\label{tab2}
\end{center}
\end{table}

The \SM $\otimes U(1)_H$ invariant Lagrangian would be
\begin{eqnarray}
  \label{EffLag}
 \mathcal{L}_{\nu}=m_{D_{ij}}\bar{L_i}\nu_{R_j} h + {M_{ij}} \bar{\nu}^c_{R_i} S_j + \\
\nonumber
M_{\sigma_{ij}} \bar{\nu}^c_{R_i} S_j\sigma + \mu_{ij}\bar{S}_iS_j +
{\mu_{\sigma_{ij}}}\bar{S}_iS_j\sigma^* + h.c.,
\end{eqnarray}
where the relevant sub-matrices are 

\begin{eqnarray}
\label{matrix}
m_D= \left[\begin{array}{ccc}
       m_a& m_b& 0\\
      m_c& m_d&0\\
       0& 0&m_e
      \end{array}
 \right],
\quad
M = \left[\begin{array}{ccc}
       0& 0& M_1\\
      0& 0&M_2\\
       0&0&0
      \end{array}
 \right], 
\\
M_\sigma = \left[\begin{array}{ccc}
       M_3& M_4& 0\\
      M_5& M_6&0\\
       0&0&M_7
      \end{array}
 \right]
\end{eqnarray}
\begin{equation}\label{matrix-2}
\mu = \left[\begin{array}{ccc}
      M_8& M_9& 0\\
      M_{10}& M_{11}&0 \\
      0&0& 0
      \end{array}
 \right],
\quad
\mu_\sigma = \left[\begin{array}{ccc}
      0&0& M_{12}\\
      0& 0&M_{13} \\
      M_{14}&M_{15}& 0
      \end{array}
 \right]
\end{equation}

One can check explicitly that the first term in Eq.~(\ref{eq:Jnunu})
is already non-diagonal and, for sufficiently low values of the
$U(1)_H$ breaking scale can induce a decay sufficiently fast as to
suppress the flux of $\phi_{\nu_{\mu}}$ to account for its non
observation of $\nu_{\mu}$ by Ice Cube.

As an additional interesting feature of this scheme, we propose an
indirect test of our neutrino decay hypothesis through the single
majoron-emitting \jznbb decay mode~\cite{Georgi:1981pg}
\begin{equation}
(A,Z) \rightarrow (A,Z+2) + 2 \, e^- + J. 
\end{equation}
The decay rate for single Majoron emission is given
by~\cite{Rodejohann:2011mu}
\begin{equation}
\Gamma^{0\nu} = \left|\vev{ g_{ee}}\right|^{2}
\left|{\mathcal M}_J \right|^2 \, G_J(Q,Z) \, , 
\label{eq:rodej}
\end{equation}
where $\left|\vev{ g_{ee}}\right|$ is an averaged coupling constant,
$G_J(Q,Z)$ accounts for the phase space factor and the nuclear matrix
element (NME) ${\mathcal M}_J$ depends on the mechanism and the
relevant nucleus.
For single Majoron emission one can use the same NMEs from the
standard $0\nu\beta\beta$ decay~\cite{Hirsch:1995in}.

\begin{figure}[!h]
\centering
\includegraphics[width=.4\textwidth]{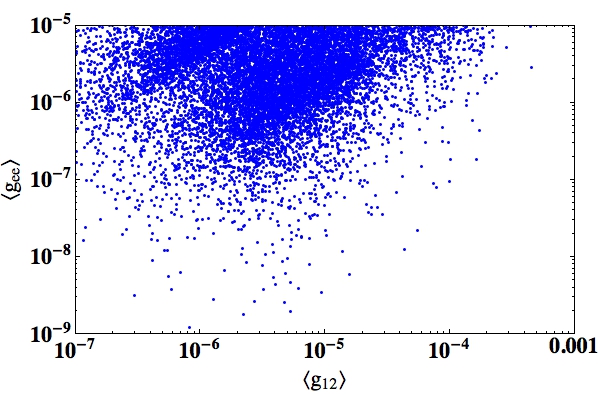} 
\includegraphics[width=.4\textwidth]{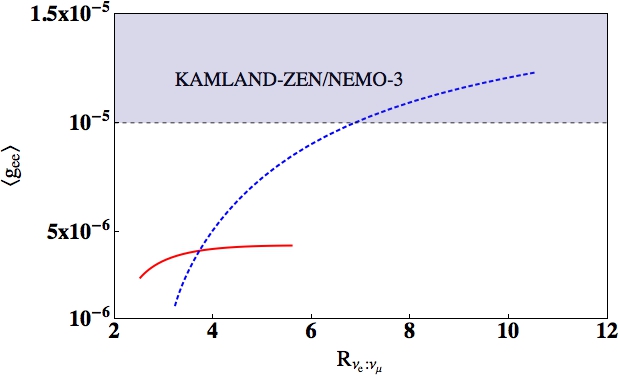} 
\caption{Left panel: Single majoron-emitting \jznbb coupling
    $\vev{ g_{ee}}$ versus $\vev{ g_{12}}$, when varying the values of
    the neutrino mass terms, for the best fit values of the neutrino
    mixing angles and $\delta = 0$. Right panel: Correlation between
    $R_{\nu_e:\nu_\mu}$ and the parameter $\vev{ g_{ee}}$ for two
    particular points of the left panel, varying in this case the
    value of $\delta$}.
\label{fig:decay-dbd}
\end{figure}

We can see from Eq.~(\ref{eq:rodej}) that the decay width for single
Majoron emission in neutrinoless double beta decay depends on the
coupling constant $g_{ee}$ and it is therefore an indirect relation
with the expression in Eq.~(\ref{eq:decayrate}) through the
coupling $g_{12}$.


Indeed if the majoron exists and its coupling to the electron neutrino
is not expected to significantly differ from the one required to
explain the muon neutrino deficit in IceCube through the neutrino
decay hypothesis, there will be a correlation between
$R_{\nu_e:\nu_\mu}$ and $\vev{g_{ee}}$. This correlation is depicted
in Fig.~\ref{fig:decay-dbd}.  We plot in the left panel the
correspondence between $g_{ee}$ and $g_{12}$ when we fix the neutrino
mixing angles at their best fit values and we consider a Dirac mass
entry at $m_D\sim$~10 GeV, $M\sim$~1 TeV $M_{\sigma}\sim$~1 TeV and
$\mu\sim$~1  keV where the $\sim$ sign takes into account order one
differences among the various flavour components of each block. 
In the right panel of the same figure, we take one the of the points shown
in the left panel and vary its CP phase from 0 to $2\pi$ in order to
obtain an estimate for the ratio $R_{\nu_e:\nu_\mu}$ relevant at
IceCube.  The dotted (blue)  curve corresponds to the case when we
  consider the second octant of the atmospheric mixing angle, particularly
  its central value $\sin^2\theta_{23}=0.61$ (we have also chosen the
  central values of the other mixing angles, $\sin^2\theta_{12} =
  0.320$ and $\sin^2\theta_{13}=0.0246$); from this case it is possible
  to see that, for example, for a coupling constant
  $\vev{g_{ee}}=4.33\times10^{-6}$ a reduction by a factor five in the
  muon flux can be obtained for an appropriate value of the $CP$ phase,
  $\delta=2.5$, while the suppression could be as high as a factor
  $10$. The solid (red) line corresponds to a point in the first
  octant ($\sin^2\theta_{23}=0.427$). Although in this particular case
  the values of the ratio, $R_{\nu_e:\nu_\mu}$ are lower than for the second 
  octant, one can still achieve an important suppression; in particular, we can see that a reduction by a factor five is again possible (for the values  
$\vev{g_{ee}}= 7.34\times10^{-6}$ and $\delta=1.6$)
This is an interesting observation, considering that currently the
``preferred'' octant is not yet uniquely determined by the neutrino
oscillation
fits~\cite{GonzalezGarcia:2012sz,Fogli:2012ua,Tortola:2012te}. 
  Moreover, we can see that even with the central values of the
  neutrino mixing angles one can obtain a suppression
  factor of five or higher, that could be sufficient to explain the
  limits reported by the IceCube 
  collaboration~\cite{Abbasi:2012zw}.

In conclusion one sees that the decay hypothesis invoked to account
for the IceCube results may be tested in the upcoming searches for the
\jznbb decay.

Work supported by MINECO grants FPA2011-22975 and MULTIDARK Consolider
CSD2009-00064, by Prometeo/2009/091 (Gen.  Valenciana), and by
EPLANET. L. D. is supported by JAE Predoctoral
fellowship. O. G. M. was supported by CONACyT grant 132197.

\bibliographystyle{frontiersinSCNS&ENG}


\end{document}